\documentclass[prl,reprint, amsmath,amssymb, aps, superscriptaddress, floatfix]{revtex4-1}
\usepackage[utf8]{inputenc}
\usepackage{graphicx}
\usepackage{dcolumn}
\usepackage{bm}
\usepackage{braket}
\usepackage{color}
\usepackage{units}
\usepackage{dsfont}
\usepackage{float}
\usepackage{hyperref}
\usepackage{xcolor}
\usepackage{ulem}
\usepackage{amsmath,amssymb,amstext,amsfonts}
\usepackage{mathtools}
\usepackage{physics}
\usepackage{amsthm}

\newcommand{\Mod}[1]{\ (\mathrm{mod}\ #1)}

\begin{document}

\title{Spatially twisted liquid-crystal devices}

\author{Alicia Sit}
\thanks{Current affiliation: National Research Council of Canada, 100 Sussex Drive, K1N 5A2, Ottawa, Ontario, Canada}
\affiliation{Nexus for Quantum Technologies, University of Ottawa, K1N 5N6, Ottawa, ON, Canada}

\author{Francesco Di Colandrea}
\affiliation{Nexus for Quantum Technologies, University of Ottawa, K1N 5N6, Ottawa, ON, Canada}

\author{Alessio D'Errico}
\affiliation{Nexus for Quantum Technologies, University of Ottawa, K1N 5N6, Ottawa, ON, Canada}

\author{Ebrahim Karimi}
\email{ekarimi@uottawa.ca}
\affiliation{Nexus for Quantum Technologies, University of Ottawa, K1N 5N6, Ottawa, ON, Canada}


\begin{abstract} 
Nematic liquid-crystal devices are a powerful tool to structure light in different degrees of freedom, both in classical and quantum regimes. Most of these devices exploit either the possibility of introducing a position-dependent phase retardation with a homogeneous alignment of the optic axis---e.g., liquid-crystal-based spatial light modulators---or conversely, with a uniform but tunable retardation and patterned optic axis, e.g., $q$-plates. The pattern is the same in the latter case on the two alignment layers. Here, a more general case is considered, wherein the front and back alignment layers are patterned differently. This creates a non-symmetric device which can exhibit different behaviours depending on the direction of beam propagation and effective phase retardation. In particular, we fabricate multi-$q$-plates by setting different topological charges on the two alignment layers. The devices have been characterized by spatially resolved Stokes polarimetry, with and without applied electric voltage, demonstrating new functionalities.
\end{abstract}

\maketitle


\section*{Introduction}
When nematic liquid crystals are placed between parallel glass plates with differing alignment directions, the bulk will twist in order to match the boundary conditions~\cite{gennes:93}. This phenomenon of a twisted nematic liquid crystal (TNLC) cell---in particular with $90^\circ$ twists---has been used extensively for the development of everyday liquid-crystal displays~\cite{shanks:82, schadt:88, yeh2009optics}. With properly chosen birefringent liquid crystals and fabrication techniques, incident linearly polarized light will rotate through the cell, following the twist structure. When a sufficiently strong voltage is applied across the cell, the twist structure disappears as the liquid crystals are aligned in the field direction, negating any polarization rotation. However, there has been limited study of the twisted cell beyond the 90$^\circ$ twist case for general polarization manipulation~\cite{gooch1975optical, ong:87}. 

On the other hand, in the context of experimental optics, spatially patterned liquid-crystal-based devices are an efficient and compact method for structuring the polarization and spatial degrees of freedom of light, but studies have been limited to symmetric elements, i.e., the front and back patterns on the alignment layers are identical. For example, $q$-plates---part of the general class of Pancharatnam--Berry phase optical elements (PBOE)---are such that the liquid-crystal layer is aligned to have a semi-integer topological charge of $q$ -- note that such a topological structure, similar to $q$-plates, may be formed by liquid crystal droplets~\cite{Brasselet:09}. This allows for the coupling of photonic spin to orbital angular momentum. $q$-plates have found applications in both classical and quantum optics~\cite{marrucci:11,rubano:19}, in particular STED microscopy~\cite{yan:15}, metrology~\cite{dambrosio:13}, high-dimensional classical~\cite{karimi:12} and quantum communication~\cite{sit:17}, and quantum simulations~\cite{cardano:15}. For the case of non-symmetric spatially patterned devices, there have been only a few implementations, including polarization converters which convert linear polarization into vector vortex modes~\cite{stalder:96}, a functionality still achievable via standard $q$-plates~\cite{cardano:12}. However, these spatially twisted elements operate with no externally applied field. Only recently, the voltage-dependent behaviour of non-symmetric devices patterned with different gratings has been observed~\cite{nys:22}.

This article aims to bridge the above gaps by investigating the behaviour of liquid crystals with the full range of possible twist angles from $-90^\circ$ to $90^\circ$, under the influence of externally applied electric fields. We first analyze the derived Jones matrix for a static TNLC cell, i.e., with no applied field, and discuss its expected behaviour in the so-called adiabatic following regime for different effective phase retardations $\Gamma$. There is a potential dual behaviour that a TNLC cell exhibits---which has not previously been reported---and it has wide-reaching implications for their spatially varying extensions. An incident circularly polarized beam may acquire three unique phase distributions from, respectively, $\Gamma=0$, $\Gamma=\pi$, and $\Gamma=0$ with reversed-plate orientation. Dual-plates (DP), as we will call them, thus promise a switch-like capability between phase distributions. Moreover, an externally applied electric field ultimately enables a transition in the effective topological charge of generated polarization patterns. Proof-of-principle spatially twisted liquid-crystal devices are fabricated and compared with the above two case studies.

\begin{figure*}[t]
	\centering
	{\includegraphics[width=\textwidth]{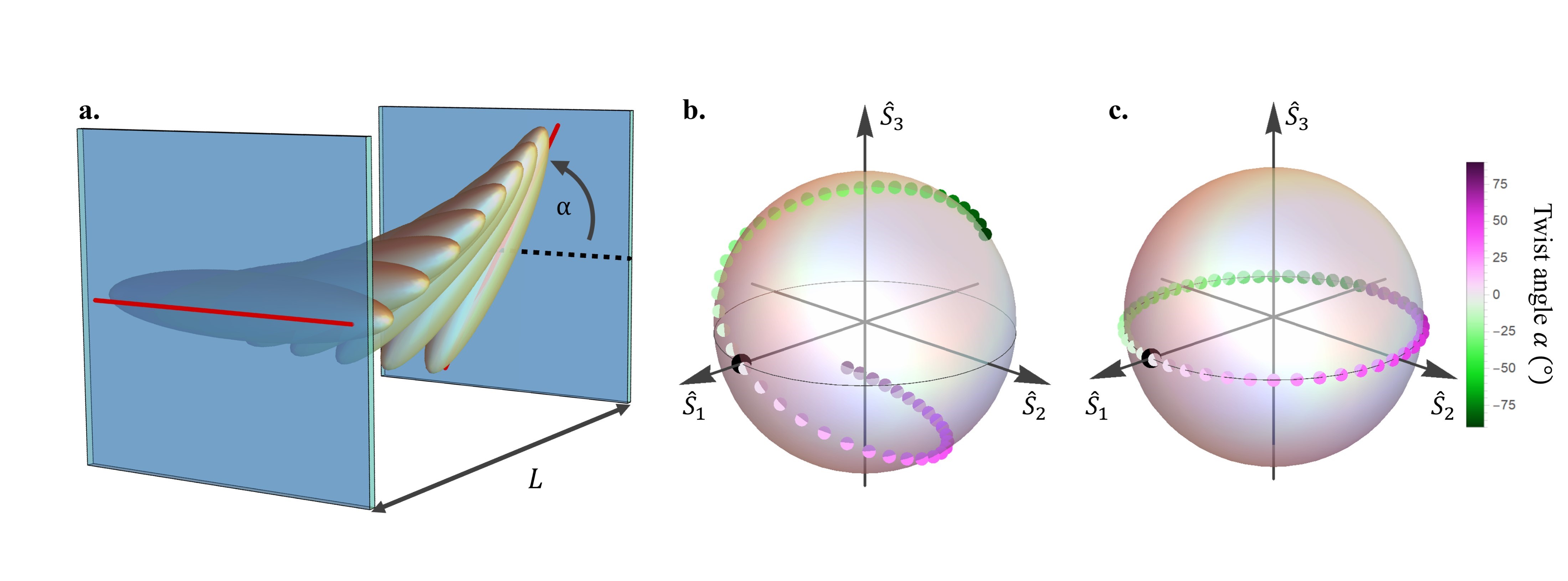}}
	\caption{\textbf{Twisted nematic liquid-crystal cells.} \textbf{a}. Illustration of liquid crystals twisting between two glass plates, uniformly aligned at $0^{\circ}$ and $\alpha$ for the front and back layers, respectively, spaced apart by a distance of $L$. The action of this configuration is shown on the Poincar\'e sphere on a horizontally polarized input state (black dot) for varying twist angles $\alpha$ between $-\pi/2$ and $+\pi/2$, with a birefringence of \textbf{b}. $\Gamma = \pi$, and \textbf{c}. $\Gamma = 1001\pi$ in the adiabatic following regime. }
	\label{fig:tnlc}
\end{figure*}

\section*{Results}
\subsection*{No applied field}
A TNLC cell of thickness $L$ without any externally applied field can be modelled as a stack of infinitely thin waveplates which are gradually twisted from one to the next. With a total twist angle of $\alpha$, and assuming strong anchoring, the twist distribution with propagation is linear, $\phi(z) = \alpha z/L$, where $z$ is the axis normal to the cell~\cite{yamauchi2005jones}. For positively uniaxial nematic liquid crystals with extraordinary and ordinary refractive indices $n_e$ and $n_o$, respectively, we can define the parameter $\Gamma= 2\pi(n_e-n_o)L/\lambda$ ($\lambda$ is the wavelength of the incident light) which corresponds to the total phase retardation for a zero-twist cell. The Jones matrix describing the TNLC configuration, shown in Fig.~\ref{fig:tnlc}-\textbf{a}, can be put in the form~\cite{yariv1983optical,yamauchi2005jones}
    \begin{eqnarray}
        \mathbf{T}_0(\alpha,\Gamma)
        &=& \mathbf{R}(-\alpha)
        \begin{bmatrix}
            \cos X -\frac{i\Gamma}{2X}\sin X & \frac{\alpha}{X}\sin X   \\
             -\frac{\alpha}{X}\sin X   & \cos X +\frac{i\Gamma}{2X}\sin X 
        \end{bmatrix} \nonumber \\
        &=:& \mathbf{R}(-\alpha)\mathbf{M}_0(\alpha,\Gamma), \label{eq:T0}
    \end{eqnarray}
with $X=\sqrt{\alpha^2+(\Gamma/2)^2}$, and the rotation matrix $\mathbf{R}(\cdot)$ is
    \begin{equation}
        \mathbf{R}(\cdot) = \begin{bmatrix}
                   \cos(\cdot) & \sin(\cdot) \\
                   -\sin(\cdot) & \cos(\cdot)
                \end{bmatrix}.
    \end{equation}
For the case where the fast axes of the front and back layers are aligned at angles $\phi_f$ and $\phi_b$, respectively, the general twisted matrix, after setting $\alpha=\phi_b-\phi_f$, becomes
    \begin{eqnarray}
        \mathbf{T}_{\phi_f}(\alpha,\Gamma)
        &=& \mathbf{R}(-\phi_f) \left[ \mathbf{R}(\phi_f-\phi_b)\mathbf{M}_0(\alpha,\Gamma) \right] \mathbf{R}(\phi_f) \nonumber \\
        &=& \mathbf{R}(-\alpha)\mathbf{M}_{\phi_f}(\alpha,\Gamma), \label{eq:tnlcr}
    \end{eqnarray}
with $\mathbf{M}_{\phi_f}(\alpha,\Gamma) = \mathbf{R}(-\phi_f)\mathbf{M}_0(\alpha,\Gamma)\mathbf{R}(\phi_f)$.

The optical action associated with Eq.~\eqref{eq:T0} can be conveniently visualized on the Poincar\'e sphere (PS). We recall that the positive (negative) points of the three principal axes $\hat{S}_1$, $\hat{S}_2$, $\hat{S}_3$ on the PS correspond, respectively, to horizontal (vertical), diagonal (anti-diagonal), and left-hand circular (right-hand circular) polarization states. Figure~\ref{fig:tnlc}-\textbf{b} demonstrates the action of a TNLC on a horizontally polarized input for twist angles between $-\pi/2$ and $\pi/2$ and a global phase retardation of $\Gamma=\pi$. In this case, the output is always elliptical, wherein the handedness is determined by the sign of $\alpha$.

Remarkably, if $\Gamma \gg \alpha$, 
then $\mathbf{M}_0(\alpha,\Gamma)$ simplifies to a waveplate $\mathbf{W}_0(\Gamma)$ with its fast axis along the horizontal: 
     \begin{equation}
        \mathbf{W}_0(\Gamma) = \begin{bmatrix}
             e^{-i\tfrac{\Gamma}{2}} & 0  \\
             0 & e^{i\tfrac{\Gamma}{2}}
        \end{bmatrix},
        \label{eqn:adiabmat}
    \end{equation}
and Eq.~\eqref{eq:tnlcr} becomes
    \begin{equation}
        \mathbf{T}_{\phi_f}(\alpha,\Gamma)
        \approx 
        \mathbf{R}(-\alpha)\mathbf{W}_{\phi_f}(\Gamma), \label{eq:adfol}
    \end{equation}
where $\mathbf{W}_{\phi_f}(\Gamma)=\mathbf{R}(-\phi_f)\mathbf{W}_{0}(\Gamma)\mathbf{R}(\phi_f)$ is a typical waveplate of retardance $\Gamma$, with fast axis oriented at $\phi_f$. Consequently, an input beam that is linearly polarized either parallel or orthogonal to $\phi_f$ will be rotated by $\alpha$ in the counterclockwise direction. This is referred to as \textit{adiabatic following}. Figure~\ref{fig:tnlc}-\textbf{c} demonstrates this behaviour for a horizontally polarized input state for a range of twist angles, wherein the output states are calculated via Eq.~\eqref{eq:T0} with $\Gamma=1001\pi$. The horizontal input---parallel to the input alignment layer---is rotated by $\alpha$ toward other linear states along the equator of the PS.

\begin{figure*}[t]
	\centering
	{\includegraphics[width=0.9\textwidth]{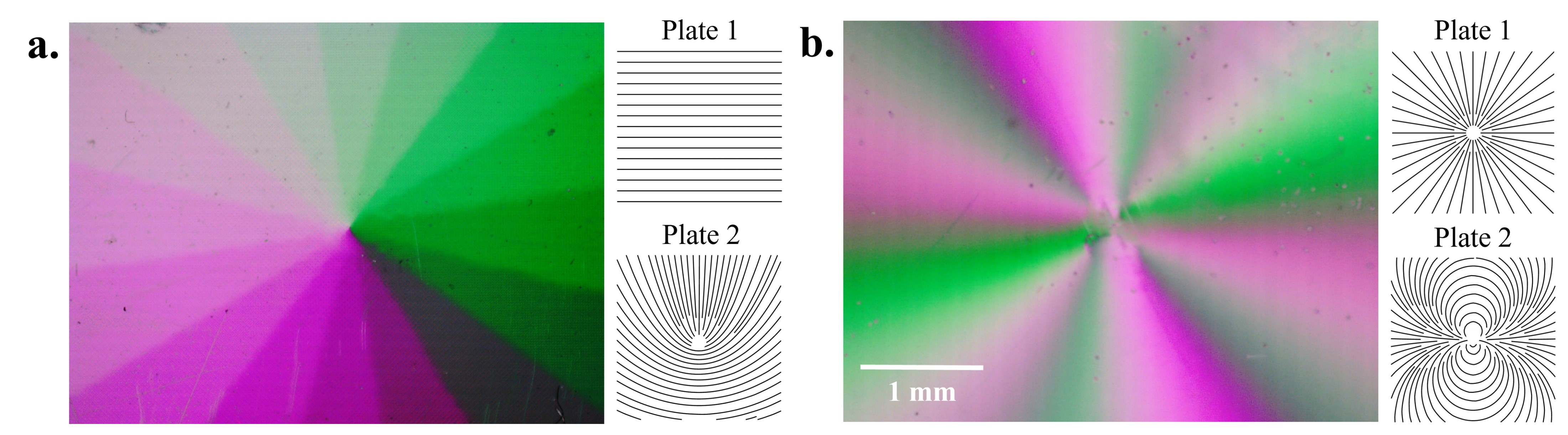}}
	\caption{\textbf{Fabricated samples.} False colour images of \textbf{a}. discretized DP(0,1/2), and \textbf{b}. DP(1,2) between crossed polarizers under a microscope illuminated with white light. The topological pattern on each glass plate is also shown. Note that the $q=1/2$ pattern is discretized into 16 slices for the patterning process.}
	\label{fig:derpPlate}
\end{figure*} 

Furthermore, the effective phase retardation is defined modulo $2\pi$ (see Eq.~\eqref{eqn:adiabmat}). Accordingly, Eq.~\eqref{eq:adfol} simplifies further for two cases: $\Gamma \Mod{2\pi}=0$ and $\Gamma\Mod{2\pi}=\pi$. For $\Gamma\Mod{2\pi}=0$, $\mathbf{W}_{\phi_f}(0)$ reduces to the identity matrix, leaving
    \begin{equation}
        \mathbf{T}_{\phi_f}(\alpha,0)  = \begin{bmatrix}
             \cos{\alpha} &  -\sin{\alpha}  \\
             \sin{\alpha} &  \cos{\alpha}
        \end{bmatrix}.  \label{eq:alpha}
    \end{equation}
Left-handed (L) and right-handed (R) circular polarizations are rotation-invariant states (up to a global phase), so explicitly acting this on the circular basis yields $\{ \ket{L},\ket{R} \} \rightarrow \{ e^{-i\alpha}\ket{L}, e^{i\alpha}\ket{R} \}$, where we have adopted bra-ket notation to denote the complex polarization spinor.
For the case where $\Gamma\Mod{2\pi}=\pi$, Eq.~\eqref{eq:adfol} simplifies to
    \begin{equation}
        \mathbf{T}_{\phi_f}(\alpha_\pi,\pi)
        = \begin{bmatrix}
             \cos{\alpha_\pi} &  \sin{\alpha_\pi}  \\
             \sin{\alpha_\pi} &  -\cos{\alpha_\pi}
        \end{bmatrix}, \label{eq:alphaPI}
    \end{equation}
with $\alpha_\pi = \phi_b+\phi_f$. This is the Jones matrix for a half-wave plate oriented at $\alpha_{\pi}/2$, yielding ${\{ \ket{L},\ket{R} \} \rightarrow \{ e^{i\alpha_\pi}\ket{R}, e^{-i\alpha_\pi}\ket{L} \}}$. 
Of course, we can also consider the case for an arbitrary $\Gamma$. In this case, Eq.~\eqref{eq:adfol} can be rewritten explicitly:
    \begin{equation}
       \mathbf{T}_{\phi_f}(\alpha,\Gamma) =   \frac{(1+e^{i\Gamma})}{2}\mathbf{T}_{\phi_f}(\alpha, 0) +\frac{(1-e^{i\Gamma})}{2}\mathbf{T}_{\phi_f}(\alpha_\pi,\pi) .
    \end{equation}
The general case is, therefore, a superposition of the cases $\Gamma\Mod{2\pi}=0$ and $\Gamma\Mod{2\pi}=\pi$, and the relative amplitudes are determined by the phase retardation.
 
In the following, we extend this analysis to spatially varying dual-plates, allowing for a spatial distribution of the fast-axis orientation of the front and back plates in the transverse plane, $\Phi_f(r,\varphi)$ and $\Phi_b(r,\varphi)$, respectively, described in cylindrical coordinates. An incident circularly polarized beam will acquire a spatially varying phase proportional to either $\alpha(r,\varphi) = \Phi_b(r,\varphi)-\Phi_f(r,\varphi)$, or $\alpha_\pi(r,\varphi) =\Phi_b(r,\varphi)+\Phi_f(r,\varphi)$, i.e., the difference or sum of the two fast axis distributions, depending on if $\Gamma\Mod{2\pi}=0$ or $\Gamma\Mod{2\pi}=\pi$. Therefore, we can use the phase retardation of a dual-plate to toggle between two different behaviours. Additionally, for a given dual-plate, if the front and back layers are reversed---i.e., the orientation of the plate is flipped, or the beam enters from the back---a distinct third phase pattern could be acquired at $\Gamma\Mod{2\pi}=0$, defined by $\alpha^{(-)}=\Phi_f - \Phi_b = -\alpha$. Of course, we could also consider the inverse problem wherein we desire a particular twist distribution; in this case, $\Phi_b = (\alpha_\pi+\alpha)/2$ and $\Phi_f = (\alpha_\pi - \alpha)/2$. 

One potential challenge that arises when dealing with the general class of non-symmetric inhomogeneous liquid-crystal plates is that nematic liquid-crystals only favourably twist between $-90^\circ$ and $90^\circ$. For example, if $\phi_f=22.5^\circ$ and $\phi_b=157.5^\circ$, then $\alpha=-45^\circ$, and not $135^\circ$, in order to achieve the lowest possible twist. At locations where $\phi_f$ and $\phi_b$ are orthogonal, there is an ambiguity as to whether $\alpha=+90^\circ$ or $-90^\circ$. This leads to discontinuities in the twist distribution, which results in $\pi$-phase jumps appearing in these locations. Consequently, a spatially varying global phase distribution will be imparted to any input beam, regardless of the birefringence setting.

An interesting example of a spatially varying dual-plate is the multi--$q$--plate.
The functionality of a $q$-plate is equivalent to that of a half-wave plate with a fast-axis distribution of $\Phi_\text{QP}(\varphi) = q\varphi + \varphi_0$, where $q$, the topological charge of the plate, is either a full or half-integer, and $\varphi_0$ is an offset angle. Its Jones matrix is identical to Eq.~\eqref{eq:alphaPI} with $\alpha_\pi=2\Phi$, and circularly polarized input will experience spin-to-orbital angular momentum coupling. For example, an input photon with a spin of $+(-)\hbar$ along the axis of propagation will gain $+(-)2q\hbar$ units of orbital angular momentum (OAM), where $\hbar$ is the reduced Planck constant, at the expense of switching the spin to be $-(+)\hbar$. We have adopted the convention that a spin of $+(-)\hbar$ corresponds to left (right) circularly polarized light. A $q$-plate can thus span the two-dimensional vector space $\{ \ket{R,2q}, \ket{L,-2q} \}$, where we have used Dirac notation with the labels corresponding to a photon's polarization and OAM, respectively.

\begin{figure*}[t]
	\centering
	{\includegraphics[width=\textwidth]{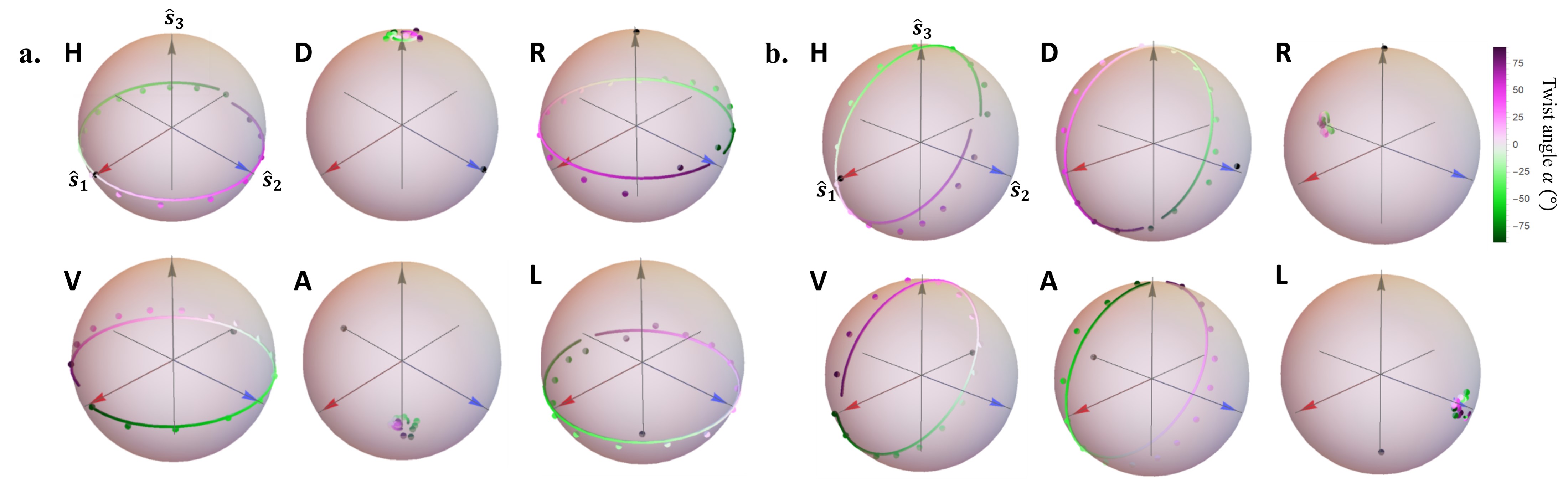}}
	\caption{\textbf{Stokes vectors reconstruction.} Reconstructed average Stokes vectors in each of the 16 slices (coloured points) for the cardinal input states (black points), and theoretical fit (line) using the TNLC Jones matrix with $\Gamma_{\text{fit}}=51.7$ for \textbf{a}. DP(0,1/2), and \textbf{b}. DP(1/2,0). The experimental Stokes vectors plotted are the average values in each slice of the discretized sample.}
	\label{fig:PSDP0HF}
\end{figure*} 

However, in order to have access to states created from a $-q$-plate---thus accessing a four-dimensional vector space---it is necessary to place a half-wave plate, or an equivalent device, to impart a $\lambda/2$ retardation, after the original $q$-plate. Additionally, it is impossible to modify the topology of a $q$-plate once this is fabricated. An appropriately patterned dual-plate is capable of performing both of these tasks without the aid of additional optical elements. For example, define the front and back distributions to be $\Phi_f(\varphi)=q_f \varphi$ and $\Phi_b(\varphi)=q_b \varphi$, respectively (we have dropped the offset angles without loss of generality). If $q_f=q_b$, we straightforwardly recover the behaviour of a regular $q=q_b$-plate so that we will assume $q_f \neq q_b$, and we get the following cases.

(1) If $q_f=0$, then $\alpha=\alpha_\pi=q_b \varphi$, and we lose the dual behaviour of the dual-plate. (2) However, if $q_b=0$, then $\alpha=-q_f \varphi$ and $\alpha_\pi=q_f \varphi$; we can thus use the phase retardation of the dual-plate to toggle between the behaviour of oppositely charged $q_f$-plates. We note that cases (1) and (2) are the same dual-plate; however, the orientation of the device---or equivalently the beam propagation---is reversed. So, while we obtain two distinct behaviours through $\alpha$ and $\alpha_\pi$, we do not gain a third behaviour since $\alpha^{(-)}=\alpha$ in case (2). (3) In general, if $|q_f|,|q_b|>0$, then $\alpha \neq \alpha_\pi$ and $\alpha^{(-)} \neq \alpha$, thus creating a multi-$q$-plate with three possible behaviours. We note that the same concept can be applied to any phase distribution. Indeed, we are not limited to $\Phi_f$ and $\Phi_b$ being of the same class of phases. In this way, we can create arbitrary dual-functionality dual-setting devices.

\begin{figure}[t]
	\centering
	{\includegraphics[width=0.47\textwidth]{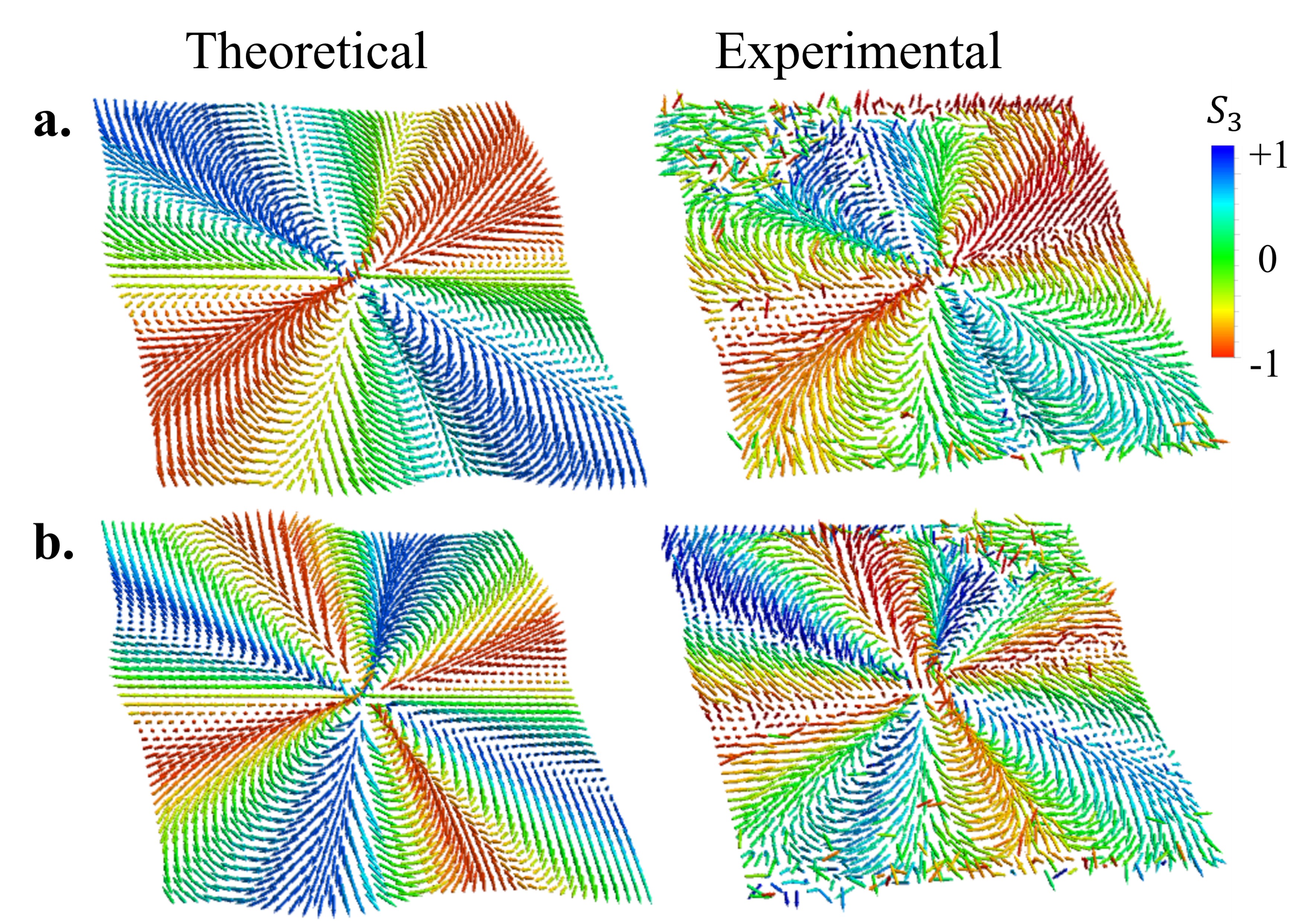}}
	\caption{\textbf{Stokes vectors reconstruction.} For horizontally polarized input light, the theoretical and experimentally reconstructed local Stokes vectors (arrows) for \textbf{a}. DP(1,2) and \textbf{b}. DP(2,1). The arrow color is a measure of the local polarization's ellipticity, with left-hand circular as red, right-hand circular as blue, and linear as green.}
	\label{fig:PSDP12}
\end{figure} 

To explore the validity of the TNLC model, several dual-plates are fabricated and characterized (see Materials and Methods). The first device is a discretized multi $q$-plate DP($q_b$=0, $q_f$=1/2), with $\sim$35~$\mu$m spacers, where DP stands for dual-plate, and the labels are the topologies of the front and back plate. Figure~\ref{fig:derpPlate}\textbf{a} shows the fabricated sample as it appears between crossed polarizers. The $q=1/2$ topology is discretized into 16 slices such that a range of twist angles from $[-90^\circ,90^\circ]$ can be efficiently characterized with enough room in each slice to average imperfections from the assemblage. The second sample fabricated and tested is a DP(1,2), shown in Fig.~\ref{fig:derpPlate}\textbf{b}. The experimentally calculated average Stokes vectors for each of the 16 slices of DP(0,1/2) with different input polarized light are plotted on the Poincar\'e sphere, along with theoretical fits (see Fig.~\ref{fig:PSDP0HF}\textbf{a}). The results for DP(1/2,0) are shown in Fig.~\ref{fig:PSDP0HF}\textbf{b}. Figures~\ref{fig:PSDP12}\textbf{a}-\textbf{b} show the locally reconstructed Stokes vectors for horizontally polarized input light on DP(1,2) and DP(2,1), respectively. A very nice agreement is observed for the two cases, despite the singularity misalignment deriving from the fabrication process.

\begin{figure*}[t]
	\centering
	{\includegraphics[width=0.95\textwidth]{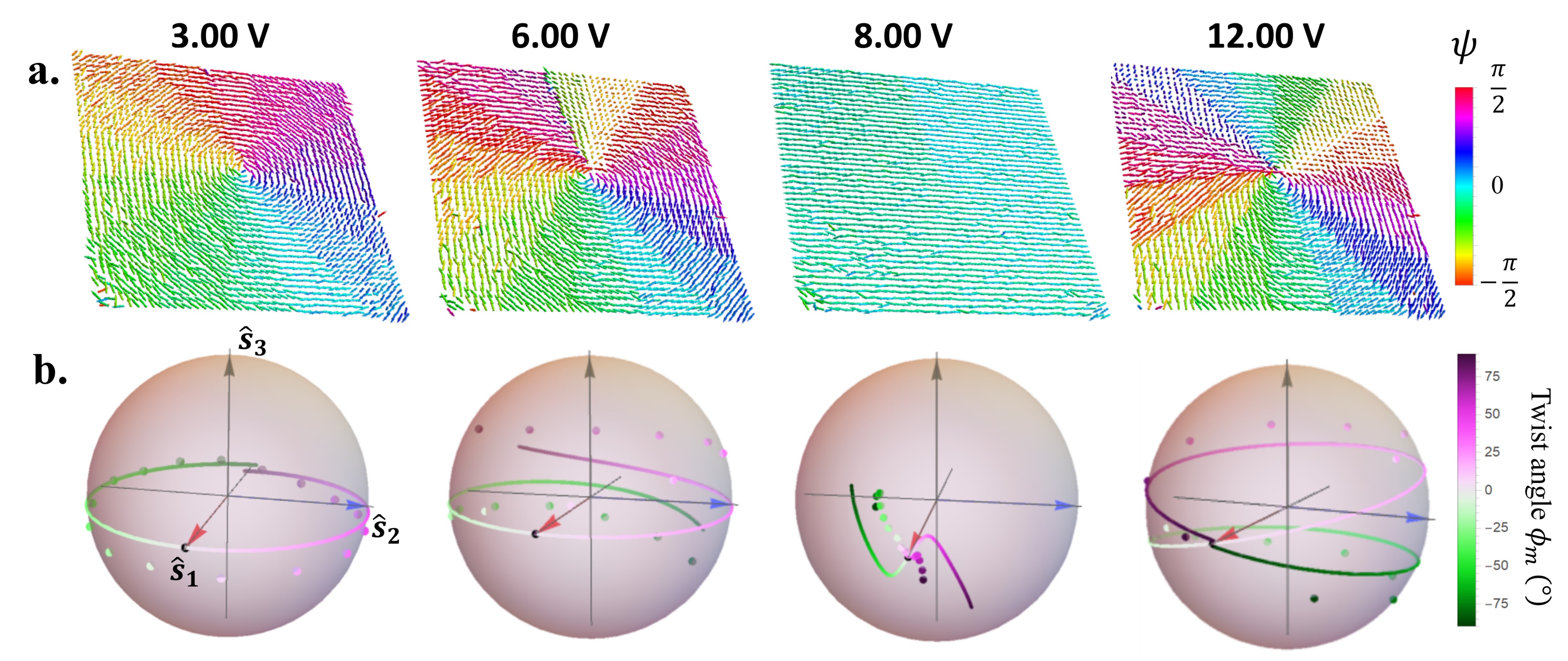}}
	\caption{\textbf{Externally applied voltage on DP(0,1/2).} \textbf{a}. Reconstructed local Stokes vectors (arrow) from a horizontally polarized input for voltage: $V_{pp}=3.00$, 6.00, 8.00, and 12.00~V. The color corresponds to the polarization ellipse angle $\psi$ with respect to the horizontal. \textbf{b}. The experimental data (dots) are the output Stokes vectors.}
	\label{fig:GA-exp-comp}
\end{figure*}

The diameters of the spacers placed between the glass plates of the cell range from 32--38~$\mu$m, with an average of 35~$\mu$m. This gives our sample an average phase retardation of $\Gamma_{avg}=51.7$ for 6CHBT liquid crystals with $\Delta n= 0.151$~\cite{lc:12}. This $\Gamma_{avg}$ is used when plotting the theoretical fits using the TNLC Jones matrix $\mathbf{T}_{\phi_f}(\alpha,\Gamma)$ of Eqs.~\eqref{eq:tnlcr} in Figs.~\ref{fig:PSDP0HF}\textbf{a}-\textbf{b}. As a measure of good fit, the average state overlap across all twist angles for a given input polarization is used.  The overlap is calculated as $|\mathbf{S}_{\text{exp}}\cdot\mathbf{S}_{\text{TNLC}}|^2$, where $\mathbf{S}_{\text{exp}}$ and $\mathbf{S}_{\text{TNLC}}$ are the experimentally-reconstructed and theoretical Stokes vectors, respectively. The uncertainties on each average Stokes vector are the standard deviations from the area used for averaging, which are less than $10\%$ for each Stokes parameter in all cases. The theoretical fits using $\Gamma_{avg}=51.7$ imitates the experimental data for all input polarizations and the two orientation cases very well, with total average overlaps of 89\% and 87\% for DP(0,1/2) and DP(1/2,0), respectively.

In the plots obtained for horizontally and vertically polarized inputs, we can observe a slight deviation from the adiabatic following regime, which predicts a mere rotation toward other linear polarizations. In this case, the adiabatic following regime could be better enforced by employing more birefringent nematic liquid crystals or larger spacers.

\subsection{Externally applied field}
When an electric voltage is applied across the DP(0,1/2) sample, the resulting effect on polarized light can be surprising. Figure~\ref{fig:GA-exp-comp}\textbf{a}~shows the reconstructed local Stokes vectors at different voltages from a horizontally polarized input state. The color is encoded to show the azimuthal angle $\psi=\arctan({S_2/S_1})$ of the Stokes vectors on the Poincar\'e sphere. We experimentally observe that the overall topological charge can be tuned from $q\sim 1/2$ to $q\sim 1$ to $q\sim 0$, and back to $q\sim 1$ as we increase the field strength. While detuning to $q=0$ is observed with standard $q$-plates, this apparent charge-doubling is never observed nor achievable. This behaviour is also not accounted for by simply varying $\Gamma$ in $\mathbf{T}_{\phi_f}(\alpha,\Gamma)$, and we must, therefore, extend our model. 

Liquid crystals may be regarded as a continuous medium with a set of elastic constants. As such, the elastic continuum theory~\cite{gennes:93} has been an excellent way to describe the influence of boundary conditions and externally applied fields---whether electric or magnetic---on these systems. Since liquid crystals are electrically polarizable, diamagnetic, and anisotropic in their electric/magnetic properties, an applied field will cause the molecules to align with the field direction. In general, the tilt and twist distributions of the liquid-crystal directors within the bulk become non-trivial, and typically non-analytical. These are determined by minimizing the Frank--Oseen free-energy density of the system~\cite{deuling:74}. The solution that arises is a coupled set of highly singular integrals to be solved simultaneously. In the high-voltage limit ($V/V_{T0}\geqslant 4$), certain approximations can be made to simplify solving these integrals around their singularities, with some analytical approximations when $V\gg V_{T0}$~\cite{preist:89}. $V_{T0}$ is the characteristic threshold voltage of the system, dependent on material parameters; here, $V_{T0}=0.966~V$ or $V_{pp}=2.733~V$. To analyze more general cases, we have opted to use a numerical minimization approach based on evolutionary methods, specifically a genetic algorithm, to also look at the response of the system for field strengths below the high-voltage limit. Remarkably, this method does not require any a-priori hypothesis on the applied field. The details of our numerical approach will be found in a separate technical paper~\cite{sit:23-technical}.

\begin{figure*}[t]
	\centering
	{\includegraphics[width=\textwidth]{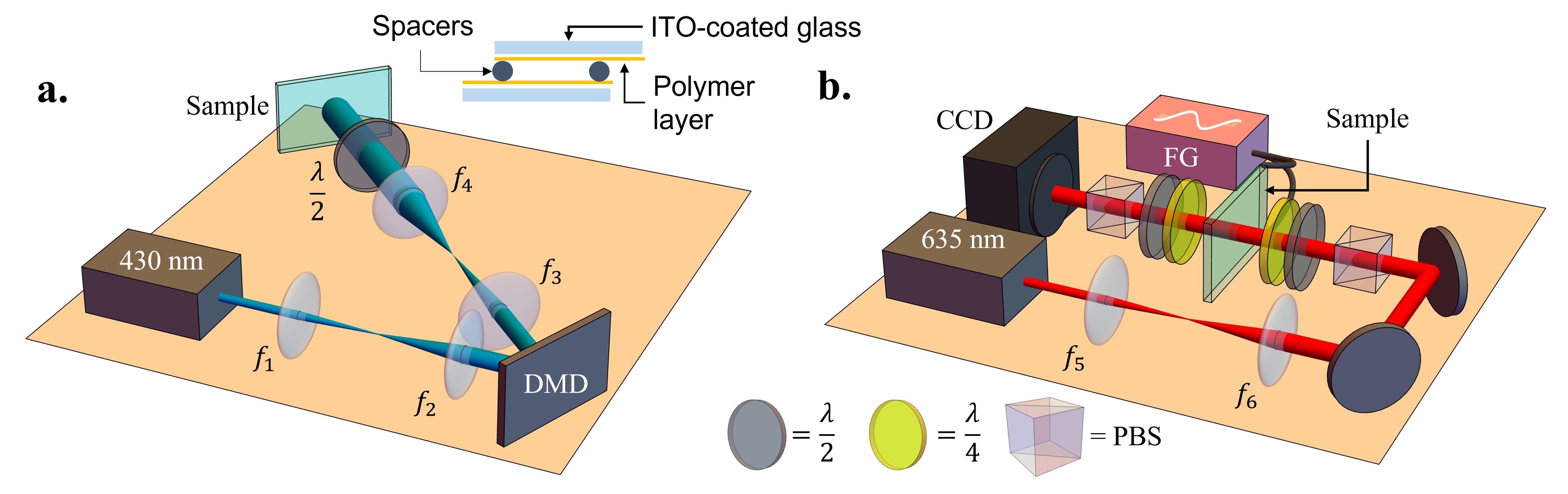}}
	\caption{\textbf{Experimental setups} \textbf{a}. Setup to pattern the glass plates for fabricating liquid-crystal devices. The inset shows the cross-section of the constructed LC cell \textbf{b}. Setup to characterize the fabricated samples. DMD = digital micromirror device, PBS = polarizing beamsplitter, FG = function generator.}
	\label{fig:setup}
\end{figure*} 

With the numerically calculated tilt and twist distributions, the overall Jones matrix $\mathbf{J}_{\phi_f}(\alpha,V)$ is derived by multiplying together the $N$ twisted liquid crystal cells of thickness $d$ using the form of Eq.~\eqref{eq:tnlcr}:
    \begin{equation} \label{eq:jm-Num}
        \mathbf{J}_{\phi_f}(\alpha,V) = \mathbf{R}(-\phi_b) \left[ \prod_{j=0}^N \mathbf{M}_0(\alpha'(z_j),\Gamma(z_j)) \right] \mathbf{R}(\phi_f),
    \end{equation}
where $\alpha'(z_j)=\phi(z_{j+1})-\phi(z_{j})$, $z_j=jd$, $\phi_f=\phi(0)$ and $\phi_b=\phi(L)$ are the front and back alignment angles, respectively, and $\alpha=\phi(L)-\phi(0)$ is the total twist angle. The phase retardation $\Gamma(z_j)$ within the $j^{\text{th}}$ slice is calculated from the tilt distribution $\theta(z)$ using the trapezoidal rule in favour of left or right Riemann sums for a better estimate,
    \begin{equation} \label{eq:gammaTrap}
       \Gamma(z_{j}) = \frac{\pi\Delta n d}{\lambda}[\cos^2\theta(z_{j+1})+\cos^2\theta(z_j) ].
   \end{equation}
A cell thickness of $L=35~\mu$m is used, with $N=100$ slices for $d=L/N$. Figure~\ref{fig:GA-exp-comp}\textbf{b} compares the experimentally reconstructed Stokes data for DP(0,1/2) and horizontally polarized input light with numerical predictions.

This model reproduces quite accurately the experimental results. This would not be possible by simply assuming a linear twist distribution and varying $\Gamma$. In particular, the non-linear twist distributions account for the peculiar phenomenon of the evolving topological charge is shown in Fig.~\ref{fig:GA-exp-comp}\textbf{a}. Further investigations will be required to determine if we achieve an actual evolution of the topological charge for the continuous version of DP(0,1/2).

\section*{Discussion and Conclusions}
We have shown the versatility of patterned twisted nematic cells, focusing on the action of dual $q$-plates on polarized light. The case in which no external field is applied has been analyzed in detail, and an analytic Jones matrix description of the dual-plates has been derived. We demonstrated that one can observe different functionalities depending on the effective retardation $\Gamma$ or the plate orientation. It must be noted that $\Gamma$ can either be fixed or tuned by means of temperature control~\cite{karimi2009efficient}. Another way of controlling the effective retardation is via an applied field. However, in the case of dual-plates, an external field introduces additional deformations in the liquid-crystal medium which affects the overall distribution of the molecular director. As a consequence, the resulting Jones matrix can be much more complicated. At the same time, this can lead to intriguing unexpected effects, for instance, the generation of polarization patterns with apparent topologies which depend on the applied voltage. Typically, a change in topology induced by a continuous parameter is associated with the existence of an intermediate critical point. However, we observe that dual $q$-plates always exhibit singular lines in the liquid-crystal orientation pattern, which are transferred to the polarization distribution of the transmitted light. Accordingly, no topological charge can be rigorously associated with the polarization distribution. Nevertheless, the observed patterns display distinguishable features, like lemon and azimuthal patterns. Dual-plates open a new avenue for optical functionalities based on liquid crystals. Here, we focused on plates with different topological structures patterned on the two alignment layers. Other possibilities will include dual Fresnel lenses, axicons, gratings, and magic windows. The use of these structures will also allow novel studies and realizations of devices introducing 3D geometric phases~\cite{slussarenko2016guiding}.

\section*{Materials and methods}
\small{The process to fabricate dual-plates is similar to that for $q$-plates~\cite{rubano:19,larocque:16}. First, we start with two glass substrates, each with a conductive layer of indium tin oxide (ITO). A drop of an azobenzene-based dye (PAAD-22, provided by BEAM Co) is deposited on top of the ITO; the sample is then spin-coated for 30~s at 4000~rpm, and baked at 120$^{\circ}$ for 5 minutes. When exposed to the light of a wavelength around the peak of the azodye's absorption spectrum, the molecules will photoalign themselves according to the light linear polarization. Here, we use a 430-nm laser. Figure ~\ref{fig:setup}\textbf{a} shows the setup to pattern the sample. In particular, a digital micro-mirror device (DLP3000 DLP\textsuperscript{\textregistered} 0.3 WVGA Series 220 DMD) is programmed to reflect a tailored intensity pattern. The resolution of the DMD is 608x684 with a micromirror pitch of 7.6~$\mu$m. A HWP can then be rotated to adjust the polarization to the required orientation. 

When fabricating $q$-plates, the two glass substrates are first glued together---with spacers between them to create a uniform cavity for the liquid crystals---and then the sample is exposed with the desired pattern. For dual-plates, each substrate is separately exposed with the front and back plate patterns, respectively; then they are glued together, taking care to overlap the two patterns as best as possible by hand. For the samples presented here, silica microspheres with diameters between 32~$\mu$m and 38~$\mu$m are used as the spacers. Note, the two substrates are glued with a lateral offset such that wires can be soldered to the conductive ITO layer; this will allow for a voltage to be applied across the sample. Finally, the nematic liquid crystals (here, 6CHBT) are injected into the cavity and the remaining sides are sealed off with glue. The completed sample is heated to $\sim$$100^\circ$C on a hot plate and cooled to room temperature once more to cement the alignment of the liquid crystals with the desired patterns. 

The setup to characterize the fabricated sample is shown in Fig.~\ref{fig:setup}\textbf{b}. A red 635-nm diode laser is used to illuminate the sample; it is expanded using a telescoping lens system to completely cover the patterned area. A polarizing beamsplitter (PBS), half-wave plate (HWP), and quarter-wave plate (QWP) are used to prepare the input polarization state. A second set of QWP, HWP, and PBS is used to project the output polarization state from the sample onto a different polarization state. Another telescoping lens system (not shown) is used to image the sample plane and shrink the beam down to fit onto a CCD camera in order to record the intensity measurement. A function generator (FG) applies a sinusoidal waveform with 4~kHz frequency and peak-to-peak voltages $V_{pp}$ between $0~V$ and $20~V$ to the prepared sample. Polarization state tomography is thus performed for the six cardinal polarizations as inputs on the sample. This consists of projecting the output polarization state onto the six cardinal states and recording the intensity using the CCD camera,  i.e., there are 6 measurements for each input state, for 36 measurements (images) total per configuration of the sample.}


\section*{Acknowledgements}
A.S. acknowledges the financial support of the Vanier graduate scholarship of the NSERC. This work was supported by the Canada Research Chairs (CRC) program and the Natural Sciences and Engineering Research Council of Canada (NSERC).

\bibliography{uo-ethesis} 
\end{document}